\documentclass{article}
\usepackage{listings}
\usepackage[margin=1.0in]{geometry}
\usepackage{float}
\usepackage{authblk}
\begin{document}

\title{Type oriented parallel programming for Exascale}
\author[1]{Nick Brown}
\affil[1]{Edinburgh Parallel Computing Centre, James Clerk Maxwell Building, Kings Buildings, Edinburgh}
\date{}
\maketitle

\begin{abstract}
{\it Whilst there have been great advances in HPC hardware and software in recent years, the languages and models that we use to program these machines have remained much more static. This is not from a lack of effort, but instead by virtue of the fact that the foundation that many programming languages are built on is not sufficient for the level of expressivity required for parallel work. The result is an implicit trade-off between programmability and performance which is made worse due to the fact that, whilst many scientific users are experts within their own fields, they are not HPC experts. 

Type oriented programming looks to address this by encoding the complexity of a language via the type system. Most of the language functionality is contained within a loosely coupled type library that can be flexibly used to control many aspects such as parallelism. Due to the high level nature of this approach there is much information available during compilation which can be used for optimisation and, in the absence of type information, the compiler can apply sensible default options thus supporting both the expert programmer and novice alike. 

We demonstrate that, at no performance or scalability penalty when running on up to 8196 cores of a Cray XE6 system, codes written in this type oriented manner provide improved programmability. The programmer is able to write simple, implicit parallel, HPC code at a high level and then explicitly tune by adding additional type information if required.}
\end{abstract}


\section{Introduction}

The difficulty of programming has been a challenge to parallel computing over the past several decades\cite{skillicorn} and as the community moves towards Exascale, where it is likely that one will take advantage of more and more cores to solve problems, then this will become more severe. It is critical that end programmers, who might not be HPC experts, can write their code in an abstract yet powerful and consistent manner if they are to take full advantage of future super computers.
\\\\
Parallel programming models largely fall into two categories: explicit parallelism and implicit parallelism. When using explicit parallelism the programmer must handle all details of data allocation, partition, distribution, communication and synchronization which is notoriously difficult. Even to the few experts explicit parallel programs, such as those using MPI, are low-level and difficult to develop, test, debug and modify.
Implicitly parallel languages, which are simpler, rely on more hidden optimization - the compiler essentially makes decisions for the end programmer. However it is not always easy to automatically make the right decisions for parallelism. For example, the initial partition and distribution of an array can significantly affect the performance of later computation and different parts of a code may require the same array to be partitioned in different directions. This is why most parallel codes currently used in real applications are hand-written and explicitly parallel - because hardware and compiler technology is not yet advanced enough to guarantee scalable and performant implicit parallelism.
\\\\
This paper proposes a trade off between explicit parallelism and implicit parallelism. Type oriented programming addresses the issue by providing the options to the end programmers to choose between explicit and implicit parallelism. The approach is to design new types governing parallelism where a programmer may choose to use these types or may choose not to use them. These types impose additional information that guides the compiler
to generate the required parallel code or conduct optimization and apply some default parallelism method when detailed information is missing. In short these types for parallelism are issued by the programmer to instruct the compiler to perform the expected actions in static analysis and code generation. They are predefined by expert HPC programmers in a type library and different target communication methods correspond to different combinations of types.
\\\\
As one moves towards Exascale, where the programmer must make efficient use of a vast amount of resource, scalability issues will force the common algorithms for solving problems to be reconsidered. The lack of suitable programming model will limit the scientific benefit that one can gain from this new generation of machine. Jacobi's algorithm is an example of a difficulty the community might face - whilst using asynchronous halo swap communication between iterations can help improve both scability and performance at large core counts it also requires far more complex code to be written. In this paper we introduce the type oriented parallel programming language, Mesham, and using this language implement versions of Jacobi's algorithm using synchronous and asynchronous communication methods.  We consider the programming benefits of expressing asynchronous Jacobi's algorithm in a type oriented manner and demonstrate performance and scalability of our approach with runs using up to 8196 cores on a Cray XE6.
\\\\
The rest of the paper is organised as follows: Section 2 reviews the background to the problem and in Section 3 we present our proposed solution. Section 4 introduces the parallel programming language, Mesham, which is used in Section 5 where we consider both the programmability and performance characteristics of our approach when applied to a case study. Section 5 draws some conclusions and considers further work.  

\section{Background}

It is widely accepted that writing parallel codes is far more complex than their sequential counterparts. Factors such as data decomposition, communication and synchronisation add additional complexity that HPC users, who are often not programming experts, can find difficult to handle. Decisions made early on, such as the method of data distribution, might be naive in the benefit of hindsight but can prove very difficult to change once the code has matured. When it comes to writing parallel codes, there is a trade off between languages. Current practice is to write parallel codes using some lower level sequential language such as C or Fortran combined with a communications library such as MPI or OpenMP. This existing approach requires the programmer to construct their code whilst considering many levels of abstraction ranging from the high level parallel system all the way down to complex low level sequential details. Other parallel programming solutions place more emphasis upon simplicity and maintainability; however the programmer can often be stuck with some default choices which impacts performance and scalability. This illustrates the fundamental trade-off in many parallel programming solutions; those solutions which provide detailed control where the programmer can tune every aspect of parallelism to achieve good performance and scalability result in complex, difficult to maintain programs and those languages which abstract the programmer sufficiently to promote simplicity but at the expense of scalability and performance.
\\\\
As we move towards Exascale, and one harnesses more and more processors to solve a problem, this challenge is likely to become even more acute. It will be very difficult to use those existing languages which allow for high levels of control, and those that make decisions and abstraction in the name of simplicity will likely scale poorly.

\section{Type oriented programming}

A large subset of languages follow the syntax \emph{Type Variablename}, such as \emph{int a} or \emph{float b}, where the programmer declares a variable. Such statements affect both the static and dynamic semantics because the compiler can perform analysis and optimisation (such as type checking) and at runtime the variable has specific attributes such as size and format. It can be thought that the programmer provides information, to the compiler, via the type. However, there is only so much that one single type can reveal, and so languages often include numerous keywords in order to allow for the programmer to express additional information. 
\\\\
Taking the C programming language as an example, in order to declare a variable \emph{m} to be a character in read only memory which is accessed many times (so the compiler might consider using a register) and might be changed externally in unpredictable ways, the code \emph{volatile register const char m} is used. Where \emph{char} is the type and \emph{volatile}, \emph{register} and \emph{const} are inbuilt language keywords. Whilst this keyword heavy approach works well for sequential languages, in the parallel programming domain there are potentially many more attributes which might need to be associated; such as where the data is located, how it is communicated and any restrictions placed upon it. Representing all this additional information via keywords would not only bloat the language, but could also introduce inconsistencies when multiple keywords were used together with potentially conflicting behaviours.
\\\\
The type oriented approach, as introduced in \cite{types}, is for the programmer to encode all variable information via the type system by combining different types together to form the overall meaning. For instance, \emph{volatile register const char m} is instead \emph{var m:Char::const::register::volatile}, where \emph{var m} declares the variable, the operator \emph{:} specifies the type and the operator \emph{::} combines types together. In this case, a \textbf{type chain} is formed by combining the types \emph{Char}, \emph{const}, \emph{register} and \emph{volatile}. Precedence is from right to left, so for example, the read only properties of the \emph{const} type override the default read \& write properties of \emph{Char}. It should be noted that some type coercions, such as \emph{Int::Char} are meaningless and so rules exist within each type to govern which combinations are allowed. It is also possible to associate arbitrary information with each type, which might be further type chains, and the type itself will give meaning to this data. To illustrate the point, if a programmer wished to suggest which register to use for variable \emph{m} then they might express \emph{register[``ax'']} where the type itself provides the context of the string argument - in this case suggesting which register to use.
\\\\
Once set in the variable declaration, the programmer can modify the semantics of a variable by changing its type later in code. Referring to the previous example, the programmer may decide to set the variable to be writable once again using the \emph{writable} type. This can be done either permanently from that point onwards using \emph{a:a::writable} (which appends the writable type to variable \emph{a}'s type chain) or for a specific expression only via \emph{(a :: writable):=99}. In such an example, if the programmer were to attempt to write to the constant variable before the writable type is applied then an error would result. In our current version of the language types must be determined at compile time but it can be very difficult or impossible for the language to support this flexible modification of types and the compiler to dynamically determine the type. For instance a conditional might rely on some user input and based upon this set the type of a variable accordingly. To ensure that types are known at compile time, typing follows lexical scoping rules and on exit from a block of code the type of a variable, if it has been modified, reverts back to what the type was when it entered that block.   
\\\\
The type oriented approach provides a number of advantages:
\begin{enumerate}
	\item {\bf Opportunities for optimisation} - Due to the programmer specifying their code in such a high level manner the compiler can obtain a much more complete view of the system and apply optimisation. The types themselves are written by domain experts which can often result in far more performant, consistent behaviour.
	\item {\bf Choice between explicit and implicit programming} - In the absence of type information the compiler will apply sensible, well documented and safe default behaviour. This can be overridden by the programmer applying additional type information to tune aspects of parallelism if required.
	\item {\bf The right person for the right job} - An HPC expert programmer can create the types using their in depth knowledge and experience. The user of the type need not understand all the underlying complexities.
	\item {\bf Changing fundamental aspects is trivial} - For example changing data decomposition only requires a change in the type rather than rewriting large portions of the code. This allows for the programmer to tune and experiment with different options once they have got their basic code working satisfactorily. 
\end{enumerate}

\section{Related work}
ZPL\cite{zpl} is an array based parallel programming language and takes advantage of the fact that common HPC applications often involve working with arrays of data with communication in ZPL being inferred by the compiler. For instance, in order to combine two arrays A and B into C, the statement \emph{C:=A + B} performs the same job as looping through each element and summing it as required in more conventional languages. Regardless of the decomposition of \emph{A}, \emph{B} and \emph{C} the compiler will issue the required communications to complete the task. As \cite{arrayproggood} illustrates, array programming and abstracting the programmer away from parallel details works work well for some problem cases. The ZPL programmer must learn new syntax and techniques but, by tying the language to array programming and making parallelism implicit, does limit it to a specific type of problem that it is suited to solving. In Mesham it is true that the programmer must also learn new concepts, but the flexibility of our approach means that it is suited to a far wider set of problems. For instance in Mesham the default behaviour of the \emph{array} type is to combine arrays \emph{A} and \emph{B} into \emph{C} when the \emph{C:=A + B} statement is encountered. Importantly, the programmer need not rely on the default implicit communication and can use additional type information to tune this aspect of their code if required.
\\\\
High Performance Fortran(HPF)\cite{hpf} is an extension to Fortran where the programmer specifies just the data partitioning and allocation, with the compiler generating communication and deciding where to place computation. Co-array Fortran (CAF)\cite{caf} is a more explicit form of parallelism where the programmer determines data partitioning, allocation, synchronisation and computation. Similarly to HPF, in CAF communication is implicit and determined by the compiler. In the type oriented the programmer is given a choice between implicit and explicit parallelism. Either they can rely on the inbuilt simple default behaviour or instead can, via types, control lower level aspects of parallelism such as the form of communication.
\\\\
HPF and CAF are examples of Partitioned Global Address Space (PGAS) languages, where the programmer considers the entire system as one global memory space which is partitioned and each block local to some process. Access to local or non local data is transparent from the programmer's point of view, but the danger is that key aspects, such as the form of data communication, are abstracted away with the programmer having no control upon these which can greatly impact performance. Numerous languages and frameworks exist to support this model but often concentrate on either performance or programmability but not both. OpenSHMEM\cite{openshmem} looks to create a standardized API for PGAS programming and implementations allow the programmer to take advantage of hardware architectures and write efficient code. The problem is that, whilst PGAS is itself a simple and abstract parallel mode, the OpenSHMEM programmer must address low level concerns such as requiring explicit barriers, synchronisations and fences. This is further compounded by the fact that OpenSHMEM is a library of functions which are called from existing languages such as C and Fortran which fundamentally requires the programmer to work in a sequential language. 
\\\\
Chapel\cite{chapel} is a recently developed PGAS language that supports the programmer expressing different abstractions of parallelism. It does this by providing implicit and explicit options and the programmer can decide which to use depending upon their code. In this manner Chapel and Mesham have some similarities but there are some critical differences in how these two languages support the implicit-explicit choice. Many of the implicit constructs in Chapel, such as reduction, are implemented via inbuilt operators contrasted to Mesham where these are types in independent libraries. In Chapel, if one where to write to a variable from multiple parallel processes at the same time then this can result in a race condition. The solution is to use a synchronisation variable instead which is expressed via the \emph{sync} keyword in the variable declaration; in the type based approach the Mesham programmer would instead use a \emph{sync} type. A benefit to our approach is in the case of multiple synchronisation constructs being used then the behaviour in a type chain, precedence being from right to left, is well defined. Whilst languages such as Chapel might disallow combinations of conflicting keywords, such as synchronisation specifiers, supporting them in a type chain allows for the programmer to mix the behaviours of different synchronisations in a predicable manner which might be desirable.
\\\\
A recent and very interesting development has been that of Co-array C++\cite{coarrayc++}. This solution integrates co-arrays into the C++ language through template libraries. The C++ programmer can then use these template libraries to annotate their source code and hence control aspects of parallelism, in this case via the co-array abstraction. The type library of Mesham has a far wider scope than the current co-array template library but it would still be possible to encode our types as a C++ template library. The benefit to this is that the programmer need not learn a new language, instead they can use familiar aspects of existing languages to write their parallel codes. However the rest of the language is fixed and whilst template libraries are well integrated into C++ if we wanted to do something with types that is not supported then this would be very difficult. Whilst the work on co-array C++ is very interesting, and might illustrate how one can express type oriented parallelism in an existing language, there has not yet been much investigation of the performance and scaling characteristics on very large core counts. The worry of integration into an existing language is that the co-array C++ programmer is fundamentally limited by current C++ compiler implementations and the optimisations that they can perform.  

\section{Mesham}

A parallel programming language, Mesham\cite{mesham}, has been created which is based around an imperative programming language with extensions to support the type oriented concept. This is used as a vehicle for exploring the type oriented paradigm and Mesham contains around fifty types in the external type library. Half of these are similar in scope to the types seen in the previous section and other types are more complex allowing for the programmer to control advanced aspects such as data decomposition and explicit communication. By default Mesham uses a partitioned global shared memory model, where memory is split up and associated with specific processes and processes can access any element. Accessing elements held in the shared memory associated with some other process results in communication which is, by default, one sided but can be overridden by using additional types.

\begin{lstlisting}[frame=lines,caption={Default one sided communication},label={lst:dftOnesided}]
var a : Int :: allocated[single[on[1]];
var b : Int :: allocated[single[on[3]]];
a:=b;
\end{lstlisting}

In listing \ref{lst:dftOnesided} two integers are allocated, \emph{a} and \emph{b} on lines 1 and 2 respectively. They each exist as a single copy in global memory and variable \emph{a} is held in the memory of process one, \emph{b} is in the memory associated with process three. At line 3 the assignment (using operator \emph{:=} in Mesham) will copy the value held in b at process 3 into the memory of process 1 to variable \emph{a}. In the absence of any further type information the communication associated with such an assignment is one sided which is guaranteed to be safe and consistent but might not be particularly performant. 

\begin{lstlisting}[frame=lines, caption={Override communication to blocking point to point},label={lst:p2p}]
var a : Int :: allocated[single[on[1]];
var b : Int :: allocated[single[on[3]]];
(a:: channel[3,1]):=b;
\end{lstlisting}

Using such default behaviour the programmer can easily get their code working and then further tune later on if required. The code in listing \ref{lst:p2p} is very similar to that of listing \ref{lst:dftOnesided} but in the assignment at line 3 the type \emph{channel} has been coerced into variable \emph{a}'s type chain. By applying additional type information the programmer has specified that, for this assignment only, instead of using the default one sided communication a point to point channel between processes three and one should be formed and used. By default this \emph{channel} type is blocking (ie. the program will only proceed once the communication has completed) but using additional type information, such as the \emph{async} type the programmer can tune this further.

Mesham exists as a research vehicle for our use of types and we have developed a compiler and associated runtime libraries to implement this language. The compiler currently translates Mesham source code into intermediate C99 code which is not intended for human reading and is then compiled by a conforming C compiler and linked against our own runtime library. The runtime library itself fulfils a number of purposes; it implements complex functionality that language types and functions require, acts as a compatibility layer over any machine specific functionality and all communications are directed through the library to underlying communication libraries and implementations. The idea behind architecting it in this manner is that the target code produced by the compiler will run on any machine with a runtime library implementation. For instance, the version of the runtime library that we used in our experiments is the Linux version which uses MPI for underlying communications. Other runtime libraries could be developed, for instance using pthreads instead of MPI or hybrid pthread-MPI communications which would be transparent to the programmer and require no code modifications apart from optional tuning.

\section{Case study: Asynchronous Jacobi}

Solving linear systems is a critical aspect of many HPC codes and whilst a number of approaches currently exist, Jacobi's algorithm is the simplest iterative solution method. Although the convergence rate of this algorithm is inferior to other, more complex methods, the fact that the only global communication required is in computing the residual makes it highly scalable. This aspect of the algorithm makes it of interest to Exascale where problems must be decomposed over a great many cores.
\\\\
In Jacobi's algorithm to solve a linear system, $Ax=b$, one starts with a trial solution $x^{0}$ and generates new solutions, iteratively, according to $x_{i}^{(k)} = \frac{1}{a_{ii}}(b_{i}-\sum_{i != j}a_{ij}x_{j}^{(k-1)})$ where $k$ is the iteration number. One stops iterating when the norm of the global residual is smaller than a specific tolerance, $\parallel Ax-b \parallel_{2} < tol $. In this paper we solve Laplace's equation, $\bigtriangledown^{2}u=0$ which can be thought of, in pseudo code, as:
\begin{lstlisting}[frame=none, numbers=none]
for all grid points
	unew(i,j,k) = 1/6 * ( u(i+1,j,k)+u(i-1,j,k)+
			u(i,j+1,k)+u(i,j-1,k)+
			u(i,j,k+1)+u(i,j,k-1) )
\end{lstlisting}

This problem uses a 7 point stencil and it can be seen that the calculation of each point only requires the nearest neighbouring point in each dimension which is a total of six in the 3D case. When solving in parallel, using a regular domain decomposition, the majority of points have neighbours which are stored locally but for those points on the partition boundary a neighbouring point is held by another process and therefore some communication is required. Instead of communicating \emph{ad\_hoc} for each required value, to minimise overhead a data swap is performed between neighbouring processes, once per iteration, to communicate all of the values required for the next iteration. This step is commonly referred to as a halo swap.
\\\\
Whilst this halo swap communication involves, at most, the number of neighbouring processes it can still be a performance bottleneck when done synchronously as each process must wait for all of their neighbours to communicate before continuing with the latest data. Alternatively the swap can be done asynchronously, where processes will not wait for communication to complete and instead use the values that they already have for the next iteration which might not be the latest but saves them from waiting. By changing the manner of communication one can obtain an algorithm which is more scalable and fault tolerant than the synchronous version. 
\\\\
Bethune at al.\cite{asyncpaper} have compared different versions of asynchronous Jacobi implemented in Fortran using MPI, OpenMP and SHMEM solving Laplace's equation with Dirichlet boundary conditions. The authors found that, for large core counts, there is some benefit to changing the mode of communication. However, a big problem is the code complexity; changing from synchronous to asynchronous communication required a rewrite of their code and the additional tricky yet uninteresting book keeping involved with asynchronous communication increased the line count by 80\%.
\\\\
Moving to Exascale many existing algorithms will need to be modified and novel techniques experimented with to run at such scale. Traditional parallel programming methods, where the complexity of this code increases sharply, is not realistic and will severely limit our ability to scale these algorithms. As previously noted, in type oriented programming the code complexity is abstracted by types and tweaking additional parameters, such as the method of communication, only requires the addition of extra type information.

\lstset{caption=Mesham synchronous Jacobi, label=lst:syncjacobi,frame=lines}
\begin{lstlisting}
var data:array[Double,nx,ny,nz]::allocated[grid[x,y,z]::single[evendist]];
var new_data:array[Double,nx,ny,nz]::allocated[grid[x,y,z]::single[evendist]];

zeroGrid(data);
var norm_b:=fillBoundaryConditions(data);

for i from 0 to maxIters {
   norm_r:=computeResidue(data);
   norm_r:=norm_r / norm_b;
   if (norm_r < threshold) break;
   
   for i from data[pid()].low to data[pid()].high {
      for j from data[pid()][i].low to data[pid()][i].high {
         for k from data[pid()][i][j].low to data[pid()][i][j].high {
            new_data[i][j][k]:=(data[i+1][j][k]+data[i-1][j][k]+
            			data[i][j+1][k]+data[i][j-1][k]+
            			data[i][j][k+1]+data[i][j][k-1]) * 1/6;
         };
      };
   };
   
   data:=new_data;
   sync data;
}
\end{lstlisting}

Listing \ref{lst:syncjacobi} contains the synchronous Jacobi's algorithm implemented in Mesham, solving Laplace's equation, to study how orienting the complexity around types can be of assistance. For brevity miscellaneous function have been omitted. At line 1 the programmer is declaring the variable \emph{data} to be a three dimensional array holding doubles which is allocated using the \emph{grid} partition type to split the array up into \emph{x} by \emph{y} by \emph{z} partitions and the \emph{single} type distributes these partitions amongst all processes evenly via the \emph{evendist} type. The \emph{evendist} type can be thought of as a cyclical distribution where partitioned blocks will be allocated to process after process and then cycle around if there are more blocks than processes. On line 2 the variable \emph{new\_data} is declared and allocated in the same manner. 
\\\\
At lines 4 and 5 we set the initial guess to be zero and then fill the boundary conditions, the function also provides the residual norm of \emph{b}. The Jacobi iteration is between lines 7 and 24; at lines 8 to 10 the solution's residual norm and then the relative norm are calculated, the later is then compared against a threshold value and terminates once the solution has converged. Lines 12 to 20 iterate through each point and set this iteration's value to be the average of the six neighbouring points at the previous iteration. Most commonly the neighbouring points will be local but for those at the boundaries it will involve communication with neighbouring processes. Line 22 copies the newly updated values into the \emph{data} array and line 23 completes any outstanding communications for that iteration. This code does not use any explicit halo or halo swapping and the required communication is done as soon as the corresponding access is issued. In this simplest case the \emph{sync} at line 23 is not actually required because, as stated, the default behaviour completes communications at line 15 as they are issued. To keep the code listings simple in this paper we have added the \emph{sync}, which does no harm, and as we now consider tuning the code and modifying the types this synchronisation point is required to complete the communications. 


\lstset{caption=Mesham synchronous Jacobi with halo, label=lst:syncjacobihalo,frame=lines,numbers=none}
\begin{lstlisting}
var data:array[Double,nx,ny,nz]::allocated[grid[halo[1], x,y,z]::single[evendist]];
\end{lstlisting}

Whilst the code in listing \ref{lst:syncjacobi} is very simple, the cost of this is that it is not optimal in performance. For each point held on a neighbouring process communication to read that value happens immediately and as a result many small communications are issued, each with some associated overhead. The programmer, once they get the code working, can then tune using additional types and listing \ref{lst:syncjacobihalo} illustrates this fact. Listing \ref{lst:syncjacobihalo} is a snippet of the declaration of variable \emph{data} with an additional \emph{halo} type, provided as an argument to the \emph{grid} partition type. The behaviour is for a halo of size \emph{n} (in this case 1) elements wide to exist around the local data and the values held in the halo correspond to that held by the appropriate neighbouring process. Upon non local data access, if this falls within the range held in the halo then that cached value is used, otherwise communication with the appropriate process is issued. The \emph{sync} keyword (line 23) directs neighbours to exchange their halo data at a predetermined point. By using this additional type it avoids many small individual messages being sent and communication happens in one large chunk between neighbours at each iteration thus lowering the overall overhead. 

\lstset{caption=Mesham asynchronous Jacobi, label=lst:asyncjacobihalo,frame=lines,numbers=none}
\begin{lstlisting}
var data:array[Double,nx,ny,nz]::allocated[grid[halo[1]::async, x,y,z]::single[evendist]];
\end{lstlisting}

By default halo swapping communication is synchronous but this can be overridden with the addition of the \emph{async} type as per listing \ref{lst:asyncjacobihalo}. As already noted, Bethune et al.\cite{asyncpaper} found that asynchronous communication required the addition of considerable tricky and uninteresting bookkeeping code in their versions which added greatly to the complexity. By using types, the Mesham programmer has abstracted away these lower level details and can concentrate on the important aspects of what their parallel code is actually doing. Our approach allows for the programmer to more readily experiment with details of parallelisation too. For instance,  Bethune et al.\cite{asyncpaper} found that by relaxing some of the locking mechanisms it made the asynchronous code faster but at the cost of a possible race condition being introduced. In their versions they had to carefully consider what needed modification and again rewrite portions of the code. Listing \ref{lst:rasyncjacobihalo} illustrates declaration of variable \emph{data} to use this racy form of asynchronous halo swapping communication and one can see that this has been achieved with the addition of the \emph{racy} type. By default the \emph{async} type generates safe asynchronous communication but the programmer, via types, has further guided the compiler to more specific behaviour.

\lstset{caption=Mesham racy asynchronous Jacobi, label=lst:rasyncjacobihalo,frame=lines,numbers=none}
\begin{lstlisting}
var data:array[Double,nx,ny,nz]::allocated[grid[halo[1]::async::racy, x,y,z]::single[evendist]];
\end{lstlisting}

The latest version of the MPI standard, MPI 3.0\cite{mpi3}, provides asynchronous collective communications to help mitigate the programmability issues discussed in this section. Instead of having to write their own asynchronous allreduce to determine the global residual, an MPI function exists to provide this functionality and these libraries are also often tuned for the specific machines and hardware which they run upon. However, the asynchronous all-reduce is only part of the story - having to issue and keep track of asynchronous halo swapping communications is still tricky and as mentioned above required a rewriting of the code. To then go from this safe asynchronous to more performant, but unsafe racy version again required some careful thinking and rewriting of the code. Whilst the new MPI standard is a step in the right direction, fundamentally the programmer would still need to address tricky and uninteresting low level details to change the form of communications. 

\subsection{Performance and scaling}
Whilst the programmability advantages of type oriented programming have been argued, in the domain of Exascale especially, a critical question to answer is whether this approach can perform and scale satisfactorily. We have carried out performance testing on HECToR, the UK National Supercomputer, a Cray XE6 and compared the versions written in Mesham against the existing Fortran90 with MPI versions detailed by Bethune et al.\cite{asyncpaper} For all test cases a local problem size of 50x50x50 elements using weak scaling has been chosen.

\begin{table}[H]
\centering
\begin{tabular}{ | c | c | c | c | }
\hline
Cores \quad&\quad Version \quad&\quad F90+MPI Runtime(s) \quad&\quad Mesham Runtime(s)\\
\hline			
512 \quad&\quad Sync \quad&\quad 132.1 \quad&\quad 150.5\\
	\quad&\quad Async \quad&\quad 146.9 \quad&\quad 145.5\\
	\quad&\quad Async racy \quad&\quad 126.4 \quad&\quad 125.7\\
\hline
2048 \quad&\quad Sync \quad&\quad 159.4 \quad&\quad 208.9\\
	\quad&\quad Async \quad&\quad 184.9 \quad&\quad 188.4\\
	\quad&\quad Async racy \quad&\quad 163.8 \quad&\quad 163.5\\
\hline
8196 \quad&\quad Sync \quad&\quad 247.1 \quad&\quad 298.7\\
	\quad&\quad Async \quad&\quad 272.5 \quad&\quad 271.8\\
	\quad&\quad Async racy \quad&\quad 264.9 \quad&\quad 265.0\\
\hline
\end{tabular}
\caption{Performance results for Jacobi implementation versions}
\label{tbl:performance}
\end{table}

Table \ref{tbl:performance} lists the performance of the existing F90+MPI implementations against versions implemented in Mesham for different core counts. One can see that, for both the safe asynchronous and racy asynchronous versions the performance is comparable between implementations and the reduced locking mechanisms in the racy version do improve the runtime. Interestingly, for the synchronous halo swap communication there is a performance gap between our Mesham code and the F90 one. Over 512 cores the Mesham version is 14\% slower, over 2048 cores 31\% slower and 8196 cores 21\% slower. The existing F90+MPI code is highly optimised, written by HPC experts and tuned for this specific problem so it is expected to perform optimally. One explanation for the synchronous version performance gap might be due to the overhead of the runtime library which is very general and superfluous for the simplest case implementation. For the more complex asynchronous versions it is likely that the overhead of the runtime library is less important compared to the F90+MPI code as this too must do bookkeeping activities. Nevertheless, it looks like there is some further optimisation which can be done on our types and runtime library to close the performance gap. 

\section{Conclusions}
In this paper we have demonstrated how types can be used, by the programmer, to provide for varying degrees of control over their code. As the community moves closer towards Exascale providing the programmer with an abstraction which allows for simplicity but also varying degrees of control in a consistent manner will be critical to take advantage of this resource. We have ensured that the type based approach and programming language Mesham scales up to core counts of 8196. When solving problems via Jacobi's algorithm with asynchronous communication, at no performance penalty, one can gain improvements in simplicity, flexibility and maintainability of their code by using types. Whilst in the synchronous case the performance testing illustrated some hit it is believed that, with additional tuning of the types and language runtime library, this can eliminated. 
\\\\
There is still plenty of further work to be done in this area such as applying the approach to real world problems on larger core counts. Currently types must be defined in external libraries which plug into the compiler; whilst these are separate and loosely coupled, ideally one would allow for the end programmer to define their own types with complex behaviour. Therefore it would be useful to further develop the theory behind types and their uses to give a firm foundation upon which the programmers can define their own types whilst maintaining benefits of programmability, performance and scalability.
\\\\
The Laplace equation that we have considered in this paper maps comfortably to our types due to the regularity of the grid. More complex real world examples often involve arbitrary grids potentially defined at runtime by some data partitioner. Types could abstract the complexities of supporting these grids, and similar to our example where we changed form of communication, the programmer would be able to change also attributes of the grids without in depth code modification being required. At the moment some attributes of the grid for instance the size and number of partitions can be determined at runtime. There is scope for further work to be done supporting irregular grids and other aspects such as the determining the grid shape dynamically.
\\\\
The Mesham language is a research vehicle to allow us to develop, test and illustrate the central ideas behind the type oriented programming paradigm. Whilst it is not realistic to expect many parallel programmers to stop using their current languages and learn an entirely new one, the concepts developed could be, in part, retrofitted to existing languages. This would provide the best of both worlds; benefits of using types as described in this paper and the programmer's familiarity with an existing language.

\end{document}